\documentclass[12pt]{article}

\usepackage{caption}
\usepackage{epsfig}
\usepackage{psfrag}
\usepackage{verbatim}
\usepackage{amsopn}
\usepackage{appendix}
\usepackage{dsfont,amsmath,amssymb,color,units,pstricks}
\usepackage{amsfonts}
\usepackage{graphicx}
\usepackage{natbib}
\usepackage{array}
\usepackage{lscape}
\usepackage{fullpage}
\usepackage{cancel}
\usepackage{url}
\usepackage{subcaption}
\usepackage{tikz}
\usepackage{ntheorem}
\usepackage{booktabs}
\usepackage{titlesec}

\allowdisplaybreaks

\newcommand{\ex}[1]{{\mathbb E}[#1]}

\newcommand{\be}{\begin{equation}}
\newcommand{\ee}{\end{equation}}
\newcommand{\bee}{\begin{equation*}}
\newcommand{\eee}{\end{equation*}}
\newcommand{\eq}[1]{Eq.~(\ref{#1})}

\theorempreskip{1\baselineskip}
\theorempostskip{1\baselineskip}
\theoremheaderfont{\bfseries\itshape}
\theoremseparator{.\smallskip}
\theoremstyle{break}
\newtheorem{theorem}{Theorem}[section]

\newtheorem{definition}[theorem]{Definition}

\newtheorem{assumption}[theorem]{Assumption}

\newtheorem{proposition}[theorem]{Proposition}

\numberwithin{equation}{section}

\titleformat{\section} 
  {\normalfont\large\bfseries}
  {(\thesection) }
  {0.5em}
  {}
  []

\titlespacing*{\section}
  {0pt}
  {2.\baselineskip}
  {1\baselineskip}

\newcommand{\myparagraph}[1]{%
  \par\vspace{\baselineskip}%
  \paragraph{\fontsize{13pt}{13pt}\selectfont #1}%
}

\begin{document}

\unitlength = 1mm
\title{Diffusive Market Impact: A Consistent Microfoundation}


\author{Julius F. Bonart}

\maketitle


\begin{abstract}
Structural price diffusivity explains many empirical regularities of market impact including the ``square-root law'' and its crossover to a linear regime for low trading rates \citep{bonart2026diffusive}. One of its central predictions is that an information-neutral trading strategy generates a diffusive impact state. We microfound this result in an economy with a trader and many arbitrageurs who progressively eliminate predictable returns. Each arbitrageur observes realized returns and a private signal of one component of the fundamental return. As aggregate private information becomes complete and under suitable convergence to a stationary limit, the trader's impact law is of the form $j=U\cdot w$, where $w$ is the innovation in the trading schedule and $U$ is causal all-pass. Impact returns are therefore white, even though no individual arbitrageur can reconstruct the underlying trading strategy. We then investigate the economic meaning of the impact phase. We argue that the $U$ which has minimum distance from the filter $L$ generating the trade flow is an especially interesting candidate: Trade flow is then minimally distorted under impact and arbitrage, and it always guarantees positive impact costs. Under a trader flow generated by a biexponential $L$ (the simplest form allowed in our model) an interesting result emerges: Impact decline is confined to a narrow band of $50\%$--$60\%$, lower than some empirical estimates and quite consistent with others. More complicated flow models can lead to different decays. Finally, we argue that in real markets, the impact phase is probably somewhat ``distorted'' in its long-term tails which allows for a full relaxation of the impact propagator. The effect is weak, long-term mean-reversion of the impact state.

\end{abstract}

\section{Introduction}

The causal response of prices to trades -- market impact -- is central to price formation and to the economics of trading. It determines how order flow enters prices and, from a practical perspective, is tantamount to a trading cost \citep{bouchaud2009priceimpact}. Market impact therefore limits the capacity of investment strategies \citep{BonelliLandierSimonThesmar2019} and is a central input to execution, portfolio construction, and backtesting.

A common empirical workflow is to measure the price response to trades, control for informational alpha, infer a primitive impact kernel, and use that kernel to predict impact under alternative trading scenarios. Two assumptions underlie this procedure.

The first concerns causality. An impulse response estimated from observational data is identified with the causal effect of trades on prices. But order flow is endogenous: trades affect returns, while expected returns and market conditions also affect trades. This is the familiar problem of causal inference \citep{pearl2009causality}, and practitioners are well aware of it \citep{webster2023handbook}.

The second concerns extrapolation. An impact law estimated under one distribution of order flow is assumed to remain valid when that distribution changes. Optimal execution models, for example, typically optimize over a family of execution schedules while holding the impact model fixed \citep{AlmgrenChriss2001}. This assumes that the market response itself does not adapt materially to the new trading regime.

In \citet{bonart2026diffusive}, I have proposed a framework that rejects this invariance assumption. 
The starting point is simple: prices are structurally diffusive. This means that in mature and liquid markets, persistent return predictability attracts arbitrage and is progressively eliminated. A sufficiently persistent change in trading behavior should therefore trigger an adaptive market response. Following this argument, it turns out that the impact law of a trader must depend on the statistics of that trader's order flow \citep{bonart2026diffusive}. 

The argument rests on the following thought experiment. Consider a trader running a long-short strategy. Compare two worlds: one in which the trader participates, and one in which she does not. The difference between the realized and counterfactual price is the trader's market impact. But if prices remain diffusive in both worlds, their difference cannot be arbitrary. Structural diffusivity imposes a covariance constraint on impact.

This constraint becomes especially strong when the trader's flow is information-neutral, meaning that it is unrelated to fundamental price innovations. This benchmark is relevant for trading driven by rebalancing, hedging, or long-horizon allocation rather than short-horizon return forecasts; see, for example, \citet{Bucci2019SlowDecay}. In this case, \citet{bonart2026diffusive} shows that impact returns must be white. Equivalently, the cumulative impact state is diffusive. 

Specifically in the linear causal setting of \citet{bonart2026diffusive}, the whitening constraint takes a particularly simple form. Let $j_t$ denote the impact return and $w_t$ the innovation in executed volume. Then
\be
\label{jUw}
j_t \propto (U\cdot w)_t \,,
\ee
where $\cdot$ denotes convolution and $U$ is a causal all-pass filter. By construction, 
\begin{equation*}
\ex{j_t j_{t+h}} \propto \delta[h] \,.
\end{equation*}

Under weak assumptions, this restriction then also determines how impact scales with time and traded volume. The empirical phenomenology of market impact, including the square-root law of metaorder impact and its crossover to a linear regime for low trading rates, is largely recovered.

Square-root impact need therefore not reflect an anomalous response mechanism \citep{BouchaudKockelkorenPotters2006}. It can arise as a by-product of diffusivity. This picture also changes the interpretation of optimal execution and price manipulation arguments. A fixed impact law may admit a cost-minimizing or apparently price-manipulating strategy, while that strategy may itself move the market outside the regime over which the impact law remains invariant. The opportunity and the model that generates it can therefore be mutually inconsistent. The impact law becomes a moving target.

\myparagraph{Main contributions.}

Two questions were left open. 
First, why should decentralized trading generate the fixed point \eq{jUw}? The result is especially puzzling because the impact law adapts to the statistics of an individual strategy even though no market participant need observe that strategy. A microfoundation should explain how such participant-specific adaptation can emerge without participant-specific identification.

Second, the special case $U=1$ is a pure surprise model: impact is proportional to the innovation in order flow. But this case implies zero expected impact costs. Positive costs therefore require a nontrivial phase. This begs the question whether the all-pass freedom always contains economically admissible phases? \citet{bonart2026diffusive} established positive expected costs for a restricted class of order-flow processes and left the general existence result open.

This paper addresses both questions.

I consider an economy with one trader and many arbitrageurs. Realized returns are public. Each arbitrageur also observes a private signal correlated with one component of the counterfactual return. No arbitrageur has enough information to reconstruct the trader's strategy. Collectively, however, their information becomes complete when every fundamental component is covered. 

Each arbitrageur uses the information available to them to eliminate a predictable component of returns. Sequentially, these interventions remove more and more forecastable structure. When the stationary limit exists, and under complete aggregate information, residual impact converges to the structural form \eq{jUw}. The fixed point predicted by structural diffusivity therefore emerges from decentralized arbitrage even though no individual arbitrageur observes the underlying strategy.

The phase $U$ has a natural economic interpretation. Part of it may be inherited from primitive impact. Part may be generated by the way prices adjust to arbitrage -- for example through delay or smoothing. Diffusivity removes predictable variation but does not uniquely determine the timing of that adjustment.

Finally, I return to impact costs and establish some new results. 
For trading schedules generated as differences of stationary inventory processes, I show that for arbitrary stationary flow statistics there exist nontrivial causal all-passes that generate strictly positive expected impact costs. Specifically, I discuss an economically plausible ``minimum distortion criterion'' under which an exogenous trader's flow is minimally perturbed by arbitrage. This is equivalent to seeking an all-pass $U^*$ that maximizes overlap with the flow filter $L$. I then show that this overlap-maximizing $U^*$ \emph{also} generates positive impact costs. Moreover, when $L$ is a biexponential kernel -- this is the simplest benchmark admitted in my framework -- impact decay generated by the corresponding $U^*$ is confined to a narrow band of $50\%$--$60\%$ of peak impact. This is result is interesting as it shows that plausible economic selection criteria can steer the all-pass freedom towards quite realistic impact dynamics.

\myparagraph{Outline.} 

This paper is organized as follows. Sec.~\ref{sec:lit} reviews the state of the art of impact modelling. Sec.~\ref{sec:review} reviews the existing diffusive impact framework introduced in \citet{bonart2026diffusive} and discussed the critical questions left open in \citet{bonart2026diffusive}. Sec.~\ref{sec:model} introduces our model: a multi-arbitrageur economy with distributed information and an exogenous trader. The section solves the arbitrageurs' forecasting problem, before Sec.~\ref{sec:adaptation} prove under certain conditions the existence of accumulation points. Such fully arbitraged markets are characterized in Sec.~\ref{sec:arbitrage}. Sec.~\ref{sec:allpass} discusses the stationary limit in which impact emerges in the form \eq{jUw}. Sec.~\ref{sec:role} discusses the economic role of the impact phase -- positive average impact costs -- and discusses an intuitive selection criterion for the impact phase under which, in an explicitly solvable model of order flow, impact is proven to decline by $50\%$ to $60\%$ of peak impact. Sec.~\ref{sec:discussion} and Sec.~\ref{sec:conclusion} recapitulate the main modelling assumptions and conclude this paper.

\myparagraph{Notation and definitions.}

Convolution is denoted $\cdot$, the discrete Kronecker delta $\delta[k]$, and the identity filter $I$.

All filters are real.

For a filter with impulse response $A=(A[k])_{k\in\mathbb Z}$, define the $Z$-transform as
\begin{equation}
A(z) := \sum_{k\in \mathbb{Z}}A[k] z^{-k} \,.
\end{equation}

A filter is causal if $A[k] = 0$ for all $k<0$. All filters are assumed to belong to the Hardy space $H^2$, equipped with the norm 
\bee
\|A\|^2 := \sum_{k\ge0} |A[k]|^2 
\eee
and inner product
\bee
\langle A,B \rangle := \sum_{k\ge0} A[k] B[k]\,.
\eee
Thus, all impulse responses are square-summable, i. e. $\|A\|^2<\infty$.

A filter $A$ is called ``minimum-phase'' if both $A$ and its inverse $A^{-1}$ are causal and stable. Equivalently, in the rational case, $A(z)$ is minimum phase if it is causally invertible and both its poles and its zeros lie strictly inside the unit circle. 

A filter $U$ is called ``all-pass'' if its frequency response has constant unit modulus, $|U(e^{i\omega})| = 1$. Equivalently, $U(z) \overline{U(1/\overline{z})} = 1$. For a rational causal stable all-pass, all its poles lie inside the unit circle. 

We work on a probability space $(\Omega,\mathcal F,\mathbb P)$. The $L^2$-norm of a real stationary stochastic process $x$ is 
\be
\|x\|^2 = \ex{|x_t|^2} \,.
\ee
All random variables are assumed to be square-integrable. Hence they are viewed as elements of the Hilbert space
\bee
L^2(\Omega,\mathcal F,\mathbb P).
\eee
equipped with the inner product
\bee
\langle X,Y\rangle=\mathbb E[X Y].
\eee

\section{Relevant literature}
\label{sec:lit}

Empirical evidence of participant-level heterogeneity in impact functions is provided by \citet{BladonMoroGalla2012}. Quite interestingly, direct evidence supporting impact invariance is also limited: To the best of our knowledge, only \citet{TothEislerBouchaud2018} compare propagators inferred from an asset manager's trades with those fitted on the rest of the market and find no significant differences. However, the order-flow statistics of the asset manager is close to that of the rest of the market. Within the diffusive impact framework of \citet{bonart2026diffusive}, this similarity limits our ability to distinguish between universal propagator and a participant-specific one. A more discriminating empirical test would compare a strategy with a genuinely different autocorrelation than the one observed in the wider market flow. To our knowledge, such a direct test is still outstanding. 

A non-universal impact law does not imply that the measured impact of metaorders -- single investment decisions executed incrementally -- must also be non-universal. In particular, a universal scaling law for measured impact, such as the square-root law \citep{SatoKanazawa2025, MaitrierLoeperKanazawaBouchaud2026}, can be compatible with participant-specific impact laws. In the framework of \citet{bonart2026diffusive}, such laws whiten individual order flow and can generate square-root metaorder impact under the relevant assumptions on order-flow statistics. 

Since Kyle's seminal paper \citep{Kyle1985}, impact models fall under two broad categories: mechanistic and information-driven. In mechanistic models impact is encoded via mechanical changes of demand and supply curves under trading pressure \citep{DonierEtAl2014}. The mechanistic interpretation emphasizes that trading is generally anonymous in modern markets; hence the collective market cannot distinguish different participants and impact is therefore represented by a universal, participant-agnostic response mechanism. Information-based approaches emphasize what prices reveal about individual trading. Square-root impact can then emerges under residual uncertainty from efficiency and fair-pricing conditions  \citep{FarmerGerigLilloWaelbroeck2013}, Bayesian learning \citep{SaddierMarsili2024}, or as a consequence of structural price diffusivity as in our framework. Here, arbitrageurs exploit the perturbations created by individual participants through their primitive, mechanical impact. Thus, \citet{bonart2026diffusive} takes mechanical impact as given but challenges that its potentially universal primitive shape remains intact under influences of arbitrage traders. 

The setup of our model draws inspiration from \citet{Hellwig1980Aggregation} and its notion of dispersed information across agents that are visible through prices. Our iterative arbitrage process is also related to \citet{HeWang1995Differential}'s paradigm of iterative generation of trading volume under private signals. More recent closely related references include \citet{Barato2013Impact} who study the statistical arbitrage opportunities created by an external flow in a Minority Game in which participants continuously adapt their strategies. \citet{MuhleKarbe2026Unified} derive power-law impact under no-arbitrage conditions in a Hawkes model; quite strikingly, they use the notion of core orders and reaction flow which may deform the primitive impact (this is also dicussed in \citet{TothEislerBouchaud2017}). This idea is central also to our framework.

\section{Review of the existing framework}
\label{sec:review}

\citet{bonart2026diffusive} derives restrictions on market impact from a single structural requirement: prices are diffusive both with and without the trader. The argument does not ask how primitive impact is generated. It asks instead which impact processes are compatible with arbitraged prices, and it finds that the answer is strongly constrained. 

The following summary is deliberately concise and informal. For details, we refer to \citet{bonart2026diffusive}.
Consider a trader whose impact is defined relative to a counterfactual market in which she does not trade. The realized return decomposes as
\be
x_t=y_t+j_t \,,
\ee
where $y_t$ is the counterfactual return in the world where the trader never trades, and $j_t$ the trader's impact which is assumed to be generated fom the trader's flow $q$. Both $x$ and $y$ are assumed to be uncorrelated noise. When $q$ is independent of $y$ so that $j$ and $y$ are uncorrelated -- this is the information-neutral benchmark considered in \citet{bonart2026diffusive} -- impact returns must therefore also be uncorrelated, and the cumulative impact state is diffusive. 

Since the bin length $\Delta t$ over which returns are defined can be varied, we may consider a different resolution, denoted by $\widetilde \Delta t$.
Choose it small enough to resolve single executions. Impact over $\widetilde \Delta t$ is then essentially mechanical. Impact diffusivity emerges quickly after a short crossover horizon. Writing $\lambda$ for the effective impact volatility per execution,
\be
\operatorname{Var}(j_t) \simeq \lambda^2 \, \mu\, \Delta t\,,
\ee
where $\mu$ is the trading rate, defined so that $\mu\Delta t$ is the number of trades within $\Delta t$.

Choose $\Delta t$ to be the duration of the metaorders in the trader's flow (assumed of fixed length for simplicity). Under buy-sell symmetry, and assuming that impact is always positive for buy and negative for sell metaorders, the first moment is proportional to the standard deviation,
\be
\ex{j_t \mid \text{buy metaorder}} = \ex{|j_t|} = Y \, \lambda \,\sqrt{\mu\, \Delta t}\,.
\ee
For Gaussian fluctuations, as an example, $Y=\sqrt{2/\pi}$. A sell order symmetrically incurs a minus sign. See \citet{bonart2026metaorder} for an explicit computation of metaorder impact under the diffusive impact model.

This expression is then readily cast into the canonical form of square-root impact,
\be
\ex{j_t | \text{buy metaorder}} \sim \text{vol} \sqrt{\text{participation ratio}}\,,
\ee
by using basic microstructure relations. 

Finally, when trader flow is generated through a linear-time-invariant filter $L$ from innovations $w$, 
\be
q = L \cdot w \,,
\ee
whiteness of impact returns requires 
\be
\label{eq:jUw}
j = cU L^{-1}\cdot q = cU \cdot w \,,
\ee
with $c = \lambda\sqrt{\mu \Delta t}$ and $U$ a causal all-pass filter. The expression \eq{eq:jUw} makes it immediately clear that the impact law -- parametrized by $U L^{-1}$ -- must depend on the statistics of $q$, i. e. \emph{is specific for each individual strategy or participant}. 

The central assumption is that continuous arbitrage keeps both the realized and the counterfactual price diffusive, i. e. that if the trader had never traded, prices would still be diffusive. In other words, diffusivity observed today is structural and not observed by chance. 

Initial reactions to this paradigm may vary -- with qualifiers ranging from "immediately compelling" to "strong assumption". 
Perhaps, the diffusivity assumption becomes more transparent under the stationary regime that \citet{bonart2026diffusive} focuses on: Markets do not have to quickly restore diffusivity when a trader enters or exits the market. In fact, the framework does not apply to a one-shot intervention, or in the case where a trading strategy enters a hitherto unexplored trading envelope that creates a structural break with its past. 

What is assumed is that the ``equilibration time'' needed for arbitrage to become effective is short enough so that impact measurements on typical observational datasets operate de facto under the stationary regime. In other words: When establishing the square-root law empirically, a long (typically years) order flow history must be analyzed, and it may be reasonable to assume that markets restore diffusivity after perturbations are detected persistently over such long horizons, even if the process is not fast. 

Under the stationary regime of trading strategies that persistently perturb the return autocorrelation, what matters for arbitrage is then not so much the signal-to-noise ratio over finite trading episodes: Any departure from price diffusivity can in principle be detected on sufficiently long horizons. A detectability criterion based on the signal-to-noise ratio (see \citet{Bouchaud2026Bonart}) is therefore not very meaningful, unless a minimum payoff expected by an arbitrageur is specified, typically a fraction of the bid-ask spread. Following this route, \citet{bonart2026diffusive} derives a ``detectability criterion'' which supports the conclusion that even small participation rates lead to whitened impact over short, intraday horizons.

\myparagraph{The critical open question.} 

The question is therefore not \emph{whether} arbitrage can whiten returns but how: It remains unclear why an optimal arbitrage intervention would lead to \eq{eq:jUw}. 

Specifically, consider the simple filtering problem where the arbitrageur observes $x_t = y_t + q_t$ under the state equation $q_t = \rho q_{t-1} + w_t$. The steady-state optimal predictor is 
\be
\widehat x_t = \rho k \sum_{\ell \ge 1} [\rho(1-k)]^{\ell-1} x_{t-\ell} \,,
\ee
with $k$ the Kalman gain. Hence, after removing predictability, the return in the arbitraged economy reads 
\be
x_t^\star = x_t - \widehat x_t = y_t + j_t \,, \qquad j_t = \sum_{\ell \ge 1} H[\ell] y_{t-\ell} + \sum_{\ell \ge 0} K[\ell] w_{t-\ell} \,, 
\ee
with the causal filters
\begin{align*}
H[\ell] &= -\rho k [\rho(1-k)]^{\ell-1} \,, \qquad \ell \ge 1 \,, \\
K[\ell] &= [\rho(1-k)]^{\ell} \,, \qquad \ell \ge 0 \,.
\end{align*}

Hence, when cast as the above linear filtering problem, the arbitrage whitens $x$ but at the price of creating ``nuisance channels'': impact also depends on past $y$. 
In other words, there is an important requirement in the derivation of \eq{jUw}: That impact is a function \emph{only} of the trader's flow \citep{bonart2026diffusive}. Abandoning it provides an escape from the all-pass restriction of impact: If $y$ and $j$ remain correlated even when $y$ and $q$ are independent, the information-neutral benchmark, in which \eq{eq:jUw} is derived, need not apply.

This assumption explains how, for instance, the Kyle model escapes \eq{eq:jUw}. Kyle's famous model~\citep{Kyle1985} can be regarded as a Kalman filter on combined order flow. Interpreting the market maker's quoting rule as the trader's causal impact, the resulting impact depends on aggregate rather than individual flow. The implication is peculiar: \emph{not} trading for a finite period may generate ``impact'' in Kyle's model. To see this, define the counterfactual as the world where the trader permanently exits the market. This would change the information conveyed by order flow and, consequently, the market maker's pricing rule. Hence, additional impact -- growing differences between actual and counterfactual prices -- can be generated even without direct activity from the trader's strategy. Such a multi-channel impact mechanism represents a significant departure from \citep{bonart2026diffusive} and indeed many other impact models (e.g. propagator and latent liquidity models).

\myparagraph{Aggregate informational completeness.}

We conclude that a consistent microfoundation of \citet{bonart2026diffusive} cannot solely rely on the mechanism above. The solution is to consider an exogenous trader in a many-arbitrageur economy, in which return whiteness emerges from repeated arbitrage interventions by agents who individually hold little information, but are collectively informed about the counterfactual $y$. This aggregate ``informational completeness'' is the central assumption in this paper. In a way that is mathematically rigorously defined, the collective arbitrageur sector ``observes'' the counterfactual $y$, although each individual arbitrageur accesses only little private information. The individual information channels may even asymptotically vanish when the number of arbitrageurs grows. 

Aggregate informational completeness may be a realistic description of modern markets: price is a deterministic function of the collective actions of all participants. Hence, the components of $y$ may be exactly ``known'', collectively, by the counterfactual market (excluding the external trader). In this sense, informational completeness is tantamount to an arbitrage sector which lies ``dense'' in the market.

In any case, informational completeness forces the nuisance $y$-channel to vanish; the exogenous trader's impact remains a sole function of her trade flow, and \citet{bonart2026diffusive}'s restrictions yield diffusive impact, \eq{eq:jUw}. See Sec.~\ref{sec:discussion} for a detailed discussion.

\section{A many-arbitrageur economy}
\label{sec:model}

We now present a suitable microfoundation. The economy contains one trader, a collection of utility maximizing arbitrageurs, and a common counterparty. All trades with the counterparty generate mechanical price impact. The trader's flow $q$ against the counterparty is taken as exogenous whereas the arbitrage trades are determined endogenously through maximizing arbitrageurs' internal utility. Absent arbitrage, returns $x$ are the sum of a fundamental return $y$ and the mechanical impact of the trader's flow.

The fundamental return is itself composed of independent components $y^{(i)}$. Each arbitrageur $i$ observes a private signal of component $i$, whereas no arbitrageur observes the primitive innovation $w$ driving trader flow. Arbitrageurs are rational in the sense that, whenever they act, they use all information available in their own filtration to form the optimal linear forecast of the next returns and trade against their predictable component. Their trading policies are therefore linear and strictly causal: today's intervention can depend on past prices and past private signals, but not on contemporaneous innovations.

\myparagraph{The trader.}

Time is discrete. Let $w$ be centered white Gaussian innovations with 
\be 
\operatorname{Var}(w_t)=1  \,. 
\ee
The trader's stationary flow is generated through a causal filter $L$,
\be q=L\cdot w \,. \ee
Here, $q_t$ denotes the trader's inventory change over the interval from $t$ to $t+1$. 
Writing $Q_t$ for her inventory,
\be Q_{t+1}-Q_t=q_t \,, \ee
we assume
\be L(z)=(1-z^{-1})\tilde L(z) \,, \qquad \text{$\tilde L$ minimum-phase} \,, \ee
so that $L(z=1)=0$ and the inventory has the stationary representation
\be Q_t=(\tilde L\cdot w)_{t-1} \,, \ee
up to a constant holding. In particular, the inventory is bounded in $L^2$.

\myparagraph{The arbitrageurs.}

Let $N$ denote the number of arbitrageurs and $M\ge N$ the number of fundamental components. Let $\{y^{(i)}\}_{i=1}^M$ be mutually independent, centered white Gaussian innovations with
\be 
\operatorname{Var}(y_t^{(i)})=\sigma_i^2>0 \,. 
\ee
The counterfactual return over the interval from $t$ to $t+1$ is
\be 
y_t=\sum_{i=1}^M y_t^{(i)} \,, \qquad \sigma_y^2:=\operatorname{Var}(y_t)=\sum_{i=1}^M\sigma_i^2 \,. 
\ee

Let $p^{(1)}$ denote the price trajectory before any arbitrageur has intervened. It can therefore be identified with the price in an economy where arbitrageurs have never been present. Let now $m$ count successive interventions, and let $i_m$ denote the arbitrageur acting at round $m$. Denote the prevailing price of that round by $p_t^{(m)}$. The price after that round is $p_t^{(m+1)}$.  
Define the return over the interval following $t$ by
\be 
x_t^{(m)}:=p_{t+1}^{(m)}-p_t^{(m)} \,. \label{eq:price-return} 
\ee

Arbitrageurs observe returns and private information on a finite interval $t = 0,\cdots T$. The forecasting information used by arbitrageur $i_m$ at $t$ is
\be 
\mathcal F_t^{(i_m, m)}:=\sigma\! \left(x^{(m)}_\tau, y_\tau^{(i_m)} : \tau=0,1,\cdots,t-1\right) \,. 
\label{eq:frozen-information} 
\ee
In other words, $i_m$ observes current and past $m$-prices, i. e. $p^{(m)}_{\tau\le t}$ at $t$. Note that the forecasting information changes with each round.

When $M=N$, the collection of private histories is informationally complete with respect to past fundamental returns. When $M>N$, the components $y^{(N+1)},\ldots,y^{(M)}$ are not privately observed by any arbitrageur. No arbitrageur directly observes the trader's flow $q$. Intermediate price trajectories and other arbitrageurs' orders are also not disclosed to $i_m$.

The finite observation window $[0, T]$ is a technical assumption that allows for a rigorous examination of convergence in this economy.

\myparagraph{The arbitrageur's intervention.}

Denote by $a_t^{(m)}$ the signed trade that arbitrageur $i_m$ executes at $t$ in round $m$. We assume that there is no terminal trade,
\be
a_T^{(m)}=0 \,.
\ee

Price impact follows the linear decomposition of \citet{AlmgrenChriss2001} into a permanent and a temporary component. Executing $a_t$ moves the contemporaneous price by $\lambda a_t$. A permanent part $\lambda' a_t$ persists at all later dates, whereas the temporary part $(\lambda-\lambda')a_t$ has decayed by $t+1$:
\be
p_t^{(m+1)}=p_t^{(m)}+\lambda a_t^{(m)}+\lambda'\sum_{\tau<t}a_\tau^{(m)} \,, \qquad 0 < \lambda'<\lambda \,. \label{eq:impact-rule}
\ee
Returns change accordingly,
\be
x_t^{(m+1)}=x_t^{(m)}+\lambda a_{t+1}^{(m)}-(\lambda-\lambda')\,a_t^{(m)} \,. \label{eq:return-intervention}
\ee
Define
\be
\rho:=\frac{\lambda}{\lambda-\lambda'} \,, \qquad z_t^{(m)}:=\sum_{h=0}^{T-1-t}\rho^h x_{t+h}^{(m)} \,, \qquad z_T^{(m)} = 0 \,. \label{eq:forward-curve}
\ee
For $0\le\lambda'<\lambda$ we have $\rho\ge1$; negative permanent impact gives $\rho<1$. Without permanent impact, $\rho=1$ and $z_t=p_T-p_t$ is the markout to the terminal price. Substituting \eqref{eq:return-intervention} into \eqref{eq:forward-curve}, the coefficient of $a_{t+k}^{(m)}$, $1\le k\le T-1-t$, is $\rho^{k-1}\big(\lambda-(\lambda-\lambda')\rho\big)=0$, and $a_T^{(m)}=0$ removes the boundary term. Hence
\be
z_t^{(m+1)}=z_t^{(m)}-(\lambda-\lambda')\,a_t^{(m)} \,. \label{eq:z-update}
\ee
The crucial property of this $z_t$ is that it is moved only by the current trade.

\myparagraph{The arbitrageur's problem.}

Per unit, the trade executed at $t$ pays the prevailing price $p_t^{(m)}+\lambda'\sum_{\tau<t}a_\tau^{(m)}$ plus the average concession $\lambda a_t/2$, so its opening execution price is
\be
p_{\text{O},t}^{(m+1)} = p_t^{(m)}+\frac{\lambda}{2} a_t^{(m)} + \lambda'\sum_{\tau<t}a_\tau^{(m)} \,.
\ee
The settlement price is the terminal price including the standalone impact of eventually unwinding $a_t$, after temporary impact has decayed:
\be
p_{\text{S},t}^{(m+1)} = p_T^{(m)}-\frac{\lambda'}{2} a_t^{(m)} + \lambda'\sum_{\tau\le T}a_\tau^{(m)} \,.
\ee
Its payoff per unit is therefore
\bee
p_T^{(m)}-p_t^{(m)}+\lambda'\sum_{\tau>t}a_\tau^{(m)}-\frac{\lambda-\lambda'}{2}\,a_t^{(m)} \,.
\eee
The permanent-impact terms from trades executed before $t$ cancel between the opening and settlement prices. The surviving first three terms capture linear forward gains whereas the last term encodes a quadratic penalty. 

We therefore model each arbitrage trade as the decision: 
\be
a_t^{(m)}=\operatorname{argmax}_{a}\left\{\left(v_t^{(m)}-\frac{\lambda-\lambda'}{2}\,a\right)a\right\}=\frac{v_t^{(m)}}{\lambda-\lambda'} \,, \label{eq:optimal-position}
\ee
where the trading signal is the conditional gain
\be
v_t^{(m)}:=\ex{p_T^{(m)}-p_t^{(m)}+\lambda'\sum_{\tau>t}a_\tau^{(m)}\mid\mathcal F_t^{(i_m,m)}} \,. \label{eq:payoff}
\ee

\begin{proposition}[Arbitrage signal]
\label{prop:signal}
If every intervention of round $m$ follows \eq{eq:optimal-position}, then
\be
v_t^{(m)}=\widehat z_t^{(m)}:=\ex{z_t^{(m)}\mid\mathcal F_t^{(i_m,m)}} \,.
\ee
\end{proposition}

To see this, suppress the round index and write $\mathbb E_t$ for the conditional expectation given $\mathcal F_t^{(i_m,m)}$. The tower property gives
\bee
v_t=\mathbb E_t\!\left[x_t+\lambda' a_{t+1}+v_{t+1}\right]=\mathbb E_t[x_t]+\rho\,\mathbb E_t[v_{t+1}] \,, \qquad v_T=0 \,,
\eee
where the second equality uses $a_{t+1}=v_{t+1}/(\lambda-\lambda')$ and $1+\lambda'/(\lambda-\lambda')=\rho$. Iterating up to $T$ yields $v_t=\mathbb E_t\big[\sum_{h=0}^{T-1-t}\rho^h x_{t+h}\big]$.

Assume also a damping factor $0\le \gamma \le 1$ that encodes incomplete execution and other frictions that may result in a departure from the ideal policy. Combining \eqref{eq:z-update} with Proposition~\ref{prop:signal} then yields
\be
z_t^{(m+1)}=z_t^{(m)} - \gamma\widehat z_t^{(m)}  \,. 
\label{eq:oneintervention}
\ee

When $\gamma = 1$, the intervention removes all predictability of $z_t$ relative to the arbitrageur's information at round $m$. However, subsequent arbitrageurs can recreate predictability relative to earlier arbitrageurs' information. This motivates a process consisting of many repeated rounds where each arbitrageur intervenes cyclically.

\section{Adaptation in the many-arbitrageur economy}
\label{sec:adaptation}

In this technical section we rigorously establish the existence of convergent subsequences of the arbitrage process in our economy. Each such cluster point is an arbitraged market in which returns are white and uncorrelated from past components of the counterfactual price. Under \emph{informational completeness} -- in which the histories of the counterfactual price components $y^{(i)}$ are known to the collective arbitrage sector -- impact satisfies an orthogonality constraint. 
This orthogonality constraint is the basis of the all-pass form of impact \eq{jUw}, as discussed in the subsequent section.

\myparagraph{Descent towards unpredictability.}

Fix $0<\gamma\le1$. For round $m$ and date $t \in [0, T]$, the conditional expectation
\bee
P_{m,t}X:=\ex{X\mid\mathcal F_t^{(i_m,m)}}
\eee
is an orthogonal projection; $P_{m,t}z_t^{(m)}$ and the forecast error $(1-P_{m,t})z_t^{(m)}$ are orthogonal. \eq{eq:oneintervention} therefore yields
\be
\|z_t^{(m+1)}\|^2=\|z_t^{(m)}\|^2 - \gamma(2-\gamma)\|\widehat z_t^{(m)}\|^2 \,. \label{eq:descent}
\ee
Each round lowers the variance of $z_t$ by a fixed fraction of the variance of the forecast that is traded away. Note that this descent concerns $z_t$ not $x_t$; the variance of returns may increase through arbitrage.

Denote $\|z_m\|^2:=\sum_{t}\|z_t^{(m)}\|^2$ the norm of the vector $z_m:=(z_t^{(m)})$. Summing \eq{eq:descent} over $t$ gives
\be
\|z_{m+1}\|^2=\|z_m\|^2 - \frac{2-\gamma}{\gamma}\|z_{m+1}-z_m\|^2 \,, \label{eq:descent-stacked}
\ee
and summing over rounds, simplifying the telescopic sum,
\be
\sum_{m\ge1}\|z_{m+1}-z_m\|^2 = \gamma^2\sum_{m\ge1}\sum_{t}\|\widehat z_t^{(m)}\|^2 \le \frac{\gamma}{2-\gamma}\|z_1\|^2<\infty \,. \label{eq:summable}
\ee
The sequence is bounded, $\|z_m\|\le\|z_1\|$, and its steps vanish, $\|z_{m+1}-z_m\|\to0$. In particular, every forecast $\widehat z_t^{(m)}$ tends to zero.

\myparagraph{Arbitraged markets on a finite window.}

The above conditions are not sufficient to show that $z^{(m)}$ is Cauchy and thus convergent (a sufficient condition would be $\widehat z^{(m)}$ to be absolutely summable rather than just square summable). For the practical purpose of this paper this does not matter: For bounded observation intervals $t = 0,1,\cdots, T$ all processes are driven by finitely many Gaussian shocks and their linear span is thus finite dimensional. Then, a convergent subsequence of $z^{(m)}$ exists whose limit has all the desired properties: Return whiteness and unpredictability with respect to all past $y^{(i)}$. 

More specifically, assume now that the model is driven by finitely many independent Gaussian shocks: $y_t^{(i)}$ for $i\le M$, $t<T$; $w_t$ for $t<T$; plus some unobserved initial state $\xi_0$, assumed to be a $d_0$-dimensional vector of mutually independent centered Gaussian variables,
\be
\operatorname{Cov}(\xi_0) = I \,.
\ee 

Let $\mathcal H$ denote their linear span. Every price and every return of every round lies in $\mathcal H$. Its dimension is $d = T(M+1) + d_0$ with $d_0$ the finite dimension of $\xi_0$. Hence, every $X\in\mathcal H$ is represented by a vector in $\mathbb{R}^d$ (with deterministic coefficients).

By \eq{eq:descent-stacked}, all $\|z^{(m)}\|$ are bounded by $\|z_1\|$. The Bolzano-Weierstrass theorem then provides a convergent subsequence. Hence, there exist a subsequence $(m_k)$ and a $z^\star$ so that 
\be
z_{m_k} \to z^\star \,.
\ee

We call $z^\star$ an arbitraged point, and call the corresponding returns $x_t^\star=z_t^\star-\rho z_{t+1}^\star$ and prices $p_t^\star$ an arbitraged market. In it, arbitrageur $i$ observes
\bee
\mathcal F_t^{i,\star}:=\sigma\!\left(x_\tau^\star,\,y_\tau^{(i)}:\tau<t\right) \,.
\eee

\begin{proposition}[Characterization of arbitraged markets]
\label{prop:arbitraged}
Every arbitraged point $z^\star$ satisfies, for all $i\le N$:
\begin{enumerate}
\item[(i)] returns are uncorrelated from all privately observed fundamental histories: 
$\operatorname{Cov}(x_t^\star, y_\tau^{(i)})=0$ for $\tau<t$ and all $1\le i \le N$;
\item[(ii)] returns are white: $\operatorname{Cov}(x_t^\star,x_\tau^\star)=0$ for $\tau\neq t$.
\end{enumerate}
\end{proposition}

The proof has three steps. \emph{(a) Projection sequences.} From \eq{eq:summable},
\be
\|z_{n+\ell} - z_n\| \le \sum_{m=n}^{n+\ell-1} \|z_{m+1} - z_m\| \to 0 \,, \qquad n \to \infty
\ee
for every $\ell$. Since $N$ is finite, this implies 
\be
\max_{\ell=1,\cdots,N}\|z_{n+\ell} - z_n\| \to 0 \,, \qquad n \to \infty \,.
\ee
Let now $z_{m_k}\to z^\star$. In particular, for any bounded sequence $0 \le \ell_k < N$,
\be
\|z_{m_k+\ell_k} - z^\star\| \to 0 \,, \qquad k \to \infty \,.
\ee

Finally, since $\widehat z^{(m_k+\ell_k)}_t = P_{m_k+\ell_k, t} z^{(m_k+\ell_k)}_t$ tends to zero by \eq{eq:summable},
\be
\label{eq:projzstarlimit}
\| P_{m_k+\ell_k,t}z_t^\star\| \le 
\| P_{m_k+\ell_k,t}(z_t^\star - z^{(m_k+\ell_k)}_t)\| + \| P_{m_k+\ell_k,t} z^{(m_k+\ell_k)}_t\| \le 
\| z^\star - z^{(m_k+\ell_k)}_t\| + \| \widehat z^{(m_k+\ell_k)}_t\| \to 0 \,. 
\ee

\emph{(b) Projection sequences with fixed arbitrageur.} Since arbitrageur $i$ acts once in every $N$ rounds, there exists a bounded sequence $\ell_k$ such that $i$ acts in all rounds $m_k + \ell_k$. In particular, $y^{(i)} = y^{(i_{m_k + \ell_k})}$ for all $k$.

\emph{(c) Arbitraged returns.} 
For $s<t$, the tower property gives
\bee
\operatorname{Cov}\left(z^\star_t, y^{(i)}_s\right) = \operatorname{Cov}\left(z^\star_t, y^{(i_{m_k + \ell_k})}_s\right) = 
\ex{P_{m_k+\ell_k,t}(z^\star_t \, y^{(i_{m_k + \ell_k})}_s)} = \ex{(P_{m_k+\ell_k,t}z^\star_t) \, y^{(i_{m_k + \ell_k})}_s} \,.
\eee
According to \eq{eq:projzstarlimit}, $P_{m_k+\ell_k,t}z^\star_t \to 0$, so
\be
\operatorname{Cov}\left(z^\star_t, y^{(i)}_s\right) = 0 \qquad \text{for $s<t$}.
\ee
Furthermore, since $x_{m_k} \to x^\star$, continuity implies
\bee
\operatorname{Cov}\left(z^\star_t, x^{(m_k)}_s\right) \to \operatorname{Cov}\left(z^\star_t, x^\star_s\right) \,,
\eee
whereas (a) with $\ell_k = 0$ implies
\bee
\operatorname{Cov}\left(z^\star_t, x^{(m_k)}_s\right) = \ex{P_{m_k,t}(z^\star_t \, x^{(m_k)}_s)} = \ex{(P_{m_k,t}z^\star_t) \, x^{(m_k)}_s} \to 0 \,;
\eee
hence
\be
\operatorname{Cov}\left(z^\star_t, x^\star_s\right) = 0 \qquad \text{for $s<t$}.
\ee 

With $x_t^\star=z_t^\star-\rho z_{t+1}^\star$, the claims follow.

\section{Arbitraged impact of an external flow}
\label{sec:arbitrage}

The trader's primitive impact is assumed to be of the same form as the arbitrageurs' impact; her trade executed between $t$ and $t+1$ raises $p_{t+1}$ by $\kappa$ and all later prices by $\kappa'$. For convenience, we also assume a different permanent-to-total impact ratio, $\lambda\kappa' - \kappa\lambda' \ne 0$; an assumption that avoids a technical difficulty in the all-pass limit. 
In terms of returns, trades are therefore filtered by the kernel
\be
G[k]=\kappa\,\delta[k]-(\kappa-\kappa')\,\delta[k-1] \,, \qquad G(z)=\kappa-(\kappa-\kappa')\,z^{-1} \,, \label{eq:primitive-kernel}
\ee
so that the unarbitraged returns are
\be
x_t^{(1)}=\sum_{i=1}^M y_t^{(i)}+\sum_{\tau = 0}^t K_1[t,\tau] w_\tau + \sum_{\ell=1}^{d_0} R_1[t, \ell]\, \xi_0[\ell] \,, \qquad K_1 = G\cdot L \,,\label{eq:initial-returns}
\ee
where $q_t = \sum_{\tau=0}^t L[t-\tau] w_\tau$ collects the in-window innovations. Returns also depend on an independent pre-window state $\xi_0$ via a finite deterministic matrix $R_1$. 

According to \eq{eq:return-intervention}, arbitrageur $i_m$ changes returns as
\be
x_t^{(m+1)}=x_t^{(m)}+\lambda a_{t+1}^{(m)}-(\lambda-\lambda')\,a_t^{(m)} \,,
\ee
with $a_T^{(m)}=0$. 
All variables are centered and jointly Gaussian, so forecasts are linear in observations, and $a_t^{(m)}$ is a linear function of the shocks dated before $t$ and of $\xi_0$. Inductively,
\be
x_t^{(m)}=\sum_{i=1}^M\sum_{\tau=0}^{t}H_{i,m}[t,\tau]\,y_\tau^{(i)}+\sum_{\tau=0}^{t}K_m[t,\tau]\,w_\tau+\sum_{\ell=1}^{d_0} R_m[t, \ell]\, \xi_0[\ell] \,. \label{eq:x-representation}
\ee
Note that the matrices $K_m$ are not generally Toeplitz, although $G$ and $L$ were assumed to be LTI. A separate, non-rigorous, LTI limit will be examined below.

\myparagraph{Invariants.}

By \eq{eq:z-update}, $z_t^{(m+1)}=z_t^{(m)}-(\lambda-\lambda')\,a_t^{(m)}$. The intervention $a_t^{(m)}$ is generated by $\xi_0$ and by innovations dated strictly before $t$. The coefficients of $z_t$ on contemporaneous $y_t^{(i)}, w_t$ are therefore invariant. Since $z_t=\sum_{\tau=t}^{T-1}\rho^{\tau-t}x_\tau$, this implies
\be
\label{eq:invariants}
\sum_{t=\tau}^{T-1}\rho^{t-\tau}K_{m}[t,\tau]=\sum_{t=\tau}^{T-1}\rho^{t-\tau}K_{1}[t,\tau] \,, \qquad \sum_{t=\tau}^{T-1}\rho^{t-\tau}H_{i,m}[t,\tau]=1 \,.
\ee

These invariants allow us to simplify the $y^{(i)}$-channels in what follows.

\myparagraph{Arbitraged returns.}

The sums in \eq{eq:invariants} are linear, hence continuous, functions of the market. They therefore hold at every arbitraged point. Each past $y_\tau^{(i)}$, $\tau<t$, $i\le N$, belongs to the information of arbitrageur $i$. By Proposition~\ref{prop:arbitraged} and \eq{eq:invariants},
\bee
0=\operatorname{Cov}\big(x_t^\star, y_\tau^{(i)}\big)=\sigma_i^2\,H_i^\star[t,\tau] \,, \qquad \tau<t \,, \quad i\le N \,.
\eee
All strictly lagged coefficients of privately observed components vanish, and \eq{eq:invariants} therefore gives $H_i^\star[t,t]=1$. Accordingly,
\be
x_t^\star=\sum_{i=1}^N y_t^{(i)}+\sum_{i=N+1}^M\sum_{\tau\le t}H_i^\star[t,\tau]\,y_\tau^{(i)}+\sum_{\tau\le t}K^\star[t,\tau]\,w_\tau+\sum_{\ell=1}^{d_0} R^\star[t, \ell]\, \xi_0[\ell] \,. \label{eq:x-window}
\ee

We now consider the informationally complete case, which yields an orthogonality constraint on the trader-noise channel.

\begin{assumption}[Aggregate informational completeness]
\label{ass:information}
Every fundamental component is privately observed by an arbitrageur:
\be M=N \,. \ee
\end{assumption}

Informational completeness is an aggregate property. Since $y_t=\sum_{i=1}^N y_t^{(i)}$, the collection of private histories determines past fundamental returns without requiring any individual arbitrageur to observe all components. Independence gives
\be 
\operatorname{Corr}^2(y_t^{(i)},y_t)=\frac{\sigma_i^2}{\sigma_y^2} \,. 
\label{eq:individual-information} \ee
This ratio describes the fraction of fundamental variance carried by component $i$. At date $t$, only its past return innovations are observed. In the symmetric case,
\bee 
\sigma_i^2=\frac{\sigma_y^2}{N} \,, \qquad \operatorname{Corr}^2(y_t^{(i)},y_t)=\frac1N \,. 
\eee
The collection of private histories can therefore be complete even when each arbitrageur observes only a small fraction of fundamental variation.

Under informational completeness,
\be
H_i^\star = I
\ee
and 
\be 
\label{eq:diffimpact}
x_t^\star = y_t + j_t^\star \,, \qquad j^\star = \Phi\left(\begin{array}{c}\xi_0 \\ w\end{array}\right)
\ee
with the causal filter $\Phi^\star = \left[R \; K^\star\right] \in \mathbb{R}^{T\times (d_0+T)}$. Moreover, whiteness of $x^\star$ and $y_t$ implies 
\be
\label{eq:Phi}
\Phi\Phi^T = K^\star(K^\star)^T + RR^T = \Lambda \qquad \text{diagonal} \,.
\ee 

We have shown, on a finite observation window, that damped sequential arbitrage reduces the variance of weighted markouts and drives their forecasts to zero. The resulting sequence admits convergent subsequences, and every such accumulation point has temporally uncorrelated returns and no residual predictability relative to any arbitrageur's information. 

If the stationary limit exists, then, under aggregate informational completeness, the limiting return decomposes as $x^\star=y+j^\star$, where $j^\star$ depends only on trader innovations and the independent initial state. The autocorrelations of $j^\star$ vanish: Impact becomes diffusive exactly as predicted in \citet{bonart2026diffusive}.

\section{The stationary all-pass limit of impact}
\label{sec:allpass}

Assume now a well-defined stationary limit away from the finite-window boundaries, in which all finite-window representations become a time-invariant system and the initial state dimension $d_0$ converges. To be clear, the previous finite-window construction does not rigorously establish the existence of such a limit which shall remain a technical assumption in this paper.

Specifically, assume $\Lambda=c^2I$, with $c>0$, and $K^\star[t,s]=k_{t-s}$ for $s\le t$ when $T\to\infty$. Write $r_t:=R[t,:]$. Since $RR^T\le c^2I$, the singular values of $R$ are bounded by $c$, and therefore
\be
\sum_{t=0}^{T-1}\|r_t\|^2 = \operatorname{tr}(R^TR) \le c^2d_0 \,,
\ee
so $\sum_{t\ge0}\|r_t\|^2<\infty$. Consequently, $r_t\to0$ and the contribution $r_t\xi_0$ vanishes for large lags.

For every lag $\ell\ge0$, the covariance identity \eq{eq:Phi} gives
\be
r_{t+\ell}r_t^T + \sum_{h=0}^t k_{h+\ell}k_h = c^2 \delta[\ell] \,.
\ee
Taking $t\to\infty$ yields
\be
\sum_{h\ge0}k_{h+\ell}k_h = c^2 \delta[\ell] \,.
\ee
These are precisely the autocorrelation identities of a white output. Hence
\be
K^\star=cU \,, \qquad |U(e^{i\omega})|=1 \quad \text{for almost every $\omega$} \,,
\ee
where $U$ is causal and all-pass.

The dimension of the initial state explains the order of $U$. Define the infinite Hankel matrix $\mathcal H[t,s]:=k_{t+s+1}$ for $t,s\ge0$. The preceding identities imply
\be
r_t r_s^T = \sum_{h\ge1}k_{t+h}k_{s+h} \,, \qquad RR^T=\mathcal H\mathcal H^T \,.
\ee
Thus $\operatorname{rank}\mathcal H=\operatorname{rank}R\le d_0$. By Kronecker's finite-rank Hankel theorem \citep{Fuhrmann1996}, the $Z$-transform $K^\star(z)$ is a rational function,
\be
K^\star(z)=\frac{p(z)}{q(z)} \,,
\ee
where $p$ and $q$ are coprime polynomials in $z$ \citep{GemignaniHankelMatrix}. Its minimal realization order is $\operatorname{rank}\mathcal H$, equal to the degree of $q$, with poles counted with multiplicity \citep{GemignaniHankelMatrix}. Since $K^\star$ is causal and its coefficients are square summable, all its poles lie strictly inside the unit circle. Hence, $U$ is a stable rational all-pass of order
\be
\operatorname{rank}\mathcal H=\operatorname{rank}R\le d_0 \,.
\ee
Equality holds when every initial-state direction affects the output, i.e. when $\operatorname{rank}R=d_0$.

If, in addition, the weighted column sums in \eq{eq:invariants} converge to those of the limiting filter, and the corresponding weighted series converge, then
\be
cU(1/\rho)=K^\star(1/\rho)=K_1(1/\rho)=G(1/\rho)L(1/\rho)=\frac{\lambda\kappa'-\kappa\lambda'}{\lambda-\lambda'}L(1/\rho) \,.
\ee
For given $L$ and $U$ with $U(1/\rho)\ne0$, this identity fixes $c$. As a side comment, the condition $L(1) = 0$ does \emph{not} imply $K^\star(1) = 0$ which would be incompatible with the all-pass result. Write $a^\text{tot} = A\cdot w$ for the total trade of the collective arbitrage sector. The impact identity then gives $K^\star(z) = K_1(z) + \left[\lambda z - (\lambda-\lambda')\right] A(z)$. Thus,
\be
cU(1) = \lambda' A(1) \,,
\ee
which is generally nonzero for nonzero $\lambda'$. 

Accordingly, under the stated assumptions, the all-pass form of impact in \citet{bonart2026diffusive}, \eq{jUw}, is recovered.

\section{The economic role of impact phase}
\label{sec:role}

In the informationally complete, arbitraged economy, whiteness restricts impact to the form $K^\star=cU$, but does not by itself determine the phase $U$. At a fixed scale $c$, this remaining degree of freedom has economically observable consequences. In particular, it determines both the correction of returns produced by the collective arbitrage sector and the impact costs borne by the trader.

\begin{definition}[Impact costs]
Let the normalized impact process be $j=U\cdot w$. We define average impact costs as minus the expected future impact mark-out of the current order,
\be
\mathcal C[U]:=-\lim_{H\to\infty}\ex{q_t\sum_{h=1}^H j_{t+h}} \,.
\ee
Since $q=L\cdot w$, $L(z)=(1-z^{-1})\tilde L(z)$, telescoping gives
\be
\mathcal C[U]=-\lim_{H\to\infty}\sum_{h=1}^H\sum_{k\ge0}L[k]U[k+h]=-\sum_{k\ge0}\tilde L[k]U[k+1] \,.
\ee
The last series converges by Cauchy--Schwarz, since $\tilde L,U\in H^2$. Thus $\mathcal C[U]$ is a bounded functional of the all-pass phase. If the impact law is $K^\star=cU$, the corresponding monetary cost is scaled by $c$.
\end{definition}

First, note that a trivial phase, $U=I$, gives
\be
\mathcal C[I]=0 \,,
\ee
because future impact innovations are uncorrelated from the current order. Hence a nontrivial phase is necessary for strictly positive impact costs.

This observation is consistent with the standard innovation view of aggregate order flow \citep{GlostenMilgrom1985, Hasbrouck1991, MadhavanRichardsonRoomans1997, KempfKorn1999, DufourEngle2000, LilloFarmer2004, BouchaudGefenPottersWyart2004, FarmerGerigLilloMike2006, BouchaudFarmerLillo2008:how_markets_digest, TarantoBormettiLillo2014}. If prices are unpredictable conditional on publicly observed aggregate order-flow information, then an order measurable with respect to that information cannot predict subsequent price changes.

But it does not imply that the mark-out of an \emph{individual} participant's flow must vanish. Individual flow need not be publicly identifiable, and positive participant-level impact costs are compatible with white returns. We therefore ask whether positive-cost phases exist for a broad class of trader-flow processes, and whether an economic mechanism may favor one of them. To be clear, this section does not establish that such a favored all-pass is attainable by the arbitrage process: while non-trivial damping factors can lead to phase creation during arbitrage, a general theorem that establishes convergence to a specific desired all-pass appears to be elusive at this stage.

\myparagraph{Minimum distortion of flow.}

A suitable economic selection criterion may be to maximize the overlap between trader flow and its impact,
\be
U^\star\in\operatorname{argmax}_{\text{$U$ causal all-pass}}\langle U,L\rangle \,.
\ee

The economic intuition is direct: The selected phase distorts as little as possible the trader flow. Since $\ex{j_t q_t} = \langle U,L\rangle$ this also maximizes the instantaneous conditional response to positive flow. Other selection criteria can be examined -- e. g. minimizing the total trade of the collective arbitrage sector or maximizing overlap with primitive impact -- but they are generally more complicated. 

Define the functional
\be
\mathcal J[U]:=\langle U,L\rangle \,.
\ee
We now examine whether this criterion also generates positive costs for the trader.

\begin{proposition}[Positive impact costs]
\label{prop:positive-impact-costs}
Suppose $L(z)=(1-z^{-1})\tilde L(z)$, where $\tilde L$ is real and minimum phase, with $L[0]>0$. Then there exists a real causal all-pass $U$ with $\mathcal C[U]>0$. In particular, every all-pass maximizer $U^*$ of $\mathcal J$ satisfies
\be
\mathcal C[U^*]\ge\frac12\mathcal J[U^*]\ge\frac12L[0]>0 \,.
\ee
\end{proposition}

Consider the deformation
\be
V_\tau(z)=\frac{1-\tau-z^{-1}}{1-(1-\tau)z^{-1}} \,, \qquad 0<\tau<1 \,.
\ee
Each $V_\tau$ is causal and all-pass, and $V_\tau\to I$ in $H^2$ as $\tau\to0^+$. Since $V_\tau\cdot U^*$ is also all-pass, maximality implies
\be
\mathcal J[V_\tau\cdot U^*]\le\mathcal J[U^*] \,.
\label{eq:deformation}
\ee
The impulse-response coefficients of the deformation are
\bee
V_\tau[0]=1-\tau \,, \qquad V_\tau[h]=-\tau(2-\tau)(1-\tau)^{h-1} \,, \qquad h\ge1 \,.
\eee
The right derivative is therefore
\be
\lim_{\tau\to0^+}\frac{\mathcal J[V_\tau\cdot U^*]-\mathcal J[U^*]}{\tau}
=-\mathcal J[U^*]-2\sum_{k\ge0}U^\star[k]\sum_{h\ge1}L[k+h]
=\mathcal J[U^*]-2\mathcal C[U^*] \,.
\ee
Together with \eq{eq:deformation}, this gives $\mathcal C[U^*]\ge\mathcal J[U^*]/2$. Finally, $I$ is an admissible all-pass, so
\be
\mathcal J[U^*]\ge\mathcal J[I]=L[0]>0 \,.
\ee

This is encouraging: Costs are guaranteed to be at least half the overlap between flow and impact returns, which is positive. 

We now examine the overlap-minimizing $U^*$ for the simplest trader filter $L$ admitted by the constraint $L(z=1)=0$: A biexponential kernel.

\myparagraph{The biexponential benchmark.}
\label{sec:biexponential}

Consider now an admissible kernel $L$ of biexponential form,
\be
L[t]=(1-z_1)z_1^t-(1-z_2)z_2^t \,, \qquad t\ge0 \,, \qquad 0<z_1<z_2<1 \,.
\ee
The filter satisfies $L[0]>0$ and $L(1)=0$. 

We now compute the all-pass maximizing its overlap with $L$. One can show that the maximizer must be a \emph{first}-order all-pass. We therefore write
\bee
U_a(z)=\frac{a-z^{-1}}{1-az^{-1}} \,.
\eee
The first-order condition in $a$ yields
\bee
\frac{(1-z_1)\sqrt{1+z_1}}{1-a^* z_1}
=\frac{(1-z_2)\sqrt{1+z_2}}{1-a^* z_2} \,.
\eee
Writing $b:=\sqrt{(1+z_1)(1+z_2)}$, its solution simplifies to
\be
a^*=\frac{z_1+z_2+b}{1+z_1z_2+b} \,, \qquad U^*=U_{a^*} \,.
\ee
Both numerator and denominator are positive, and their difference is $(1-z_1)(1-z_2)>0$. Hence $0<a^*<1$.

Since $L(1)=0$, the costs can also be written as
\bee
\mathcal C[U]=\sum_{k\ge0}U[k]\sum_{\ell\ge0}L[k+\ell] \,.
\eee
For the biexponential kernel, $\sum_{\ell\ge0}L[k+\ell]=z_1^k-z_2^k$. Consequently,
\be
\mathcal C[U^*]=U^*(1/z_1)-U^*(1/z_2)
=\frac{(z_2-z_1)(1-(a^*)^2)}{(1-a^* z_1)(1-a^* z_2)} \,.
\ee

\begin{proposition}[Impact decay for a biexponential flow]
\label{prop:biexponential-decay}
For the biexponential kernel, the phase selected by the maximum overlap principle satisfies
\be
\frac12<\frac{\mathcal C[U^*]}{\mathcal J[U^*]}
=\frac{2+b}{3+z_1z_2+2b}<\frac35 \,, \qquad b=\sqrt{(1+z_1)(1+z_2)} \,.
\ee
The mean cumulative impact therefore declines by $50\%$ to $60\%$ of its contemporaneous peak, leaving between $40\%$ and $50\%$ at longer horizons.
\end{proposition}

The expression for the ratio follows by substituting $a^*$ into $\mathcal C$ and $\mathcal J$. The lower bound is equivalent to $z_1z_2<1$, while the upper bound is equivalent to $b+3z_1z_2>1$. Both inequalities are strict. The upper bound is approached when $z_1,z_2\to0$, and the lower bound when $z_1,z_2\to1$.

To interpret the ratio, note that
\be
\ex{j_t\mid q_t}=\frac{\ex{j_tq_t}}{\operatorname{Var}(q_t)}q_t
=\frac{\mathcal J[U^*]}{\operatorname{Var}(q_t)}q_t \,.
\ee
Moreover, the first-order all-pass yields the conditional cumulative response
\be
\ex{\sum_{h=0}^H j_{t+h}\mid q_t}
=\frac{\mathcal J[U^*]-\mathcal C[U^*](1-(a^*)^H)}
{\operatorname{Var}(q_t)}q_t \,, \qquad H\ge0 \,.
\ee
For a buy order, this response is maximal at $H=0$ and decreases monotonically towards its long-horizon plateau. Thus $\mathcal J$ controls the contemporaneous peak, while $\mathcal C$ controls the subsequent decay. 

Interpret each bin as individual autocorrelated trading decision (with potentially many child orders). Then, the above result is a statement about impact decline after the completion of the (fixed-duration) metaorder \emph{embedded in their autocorrelated order flow}. The prediction somewhat lower than the $\approx 2/3$ permanent~\citep{moro2009markethidden, gomes2015market, ZarinelliEtAl2015, Said2017MarketImpact,DonierBonart2015Million} or next-day impact level~\citep{brokmann2015slow, Bucci2019SlowDecay} reported in many studies. After applying a deconvolution procedure, \citet{Bucci2019SlowDecay} find a permanent impact level of roughly $0.42$ in units of the metaorder peak impact.

\paragraph{A distorted phase with full relaxation and bounded impact.}

All rational real all-passes are constrained to
\be
|U[t]| \le 1 \,, \qquad U(1) = \sum_{t\ge0} U[t] = \pm 1 \,.
\ee
Accordingly, impact of an isolated innovation impulse tends to a level that exceeds in magnitude the temporary impact rather than relax. While counterintuitive, this is not incompatible with the framework in which isolated impulses do not naturally occur: Permanent impact arises from a convolution of impacts and can therefore be well-behaved for naturally \emph{embedded} trading events, as is indeed the case for the example in Sec.~\ref{sec:biexponential}. More generally, for every $0\le \alpha <1$ and every $L$ there exists an all-pass that produces a permanent impact equal to a fraction $\alpha$ of contemporaneous impact. 

However, it is also clear that real prices cease to be diffusive on long time scales when trend and mean-reversion effects set in. \citet{bonart2026diffusive} points out that impact cannot remain diffusive over long horizons. 

Consider the following all-passes / low-pass combinations:
\begin{align}
V &= U - U(1)H_\rho \,, \qquad H_\rho(z) = \frac{(1-\rho)z^{-1}}{1-\rho z^{-1}} \,, \qquad 0<\rho<1 \,. \\
W &= (1-H_\rho)\cdot U \,.
\end{align}
By construction, $W(1) = V(1) = 0$, i.e. full relaxation is restored. The distortion vanishes uniformly over lags as $\rho \to 1$ since
\bee
\| V - U\|^2 = \| W - U\|^2 = \frac{1-\rho}{1+\rho} \to 0 \,.
\eee
The crossover horizon is $\tau_c = -1/\log \rho$. 

The resulting impact autocorrelation is dominated by $U$ for small lags, and impact returns are approximately white. But the exact identity $\sum_h R_{jj}[h] = V(1)^2 = 0$ also implies that, on avergage,
\bee
R_{jj}[h] < 0 \,, \qquad h \gg \tau_c \,.
\eee
Hence, impact tends to mean-revert on long horizons. 

The distorted all-pass formulation has the additional advantage of bounding the variance of the initially diffusive impact state: the mispricing caused by a trader's flow saturates at a variance of order $\frac{1}{2}(1-\rho)^{-1}$ in units of the per-period impact variance, instead of growing indefinitely.

\section{Discussion}
\label{sec:discussion}

Informational completeness is a substantial assumption in this paper. The essential role played by private information in price formation is recognized in the economics literature, but it remains almost anathema to the econophysics school of finance.
Let us therefore critically examine the assumption. 

Aggregate informational completeness does not imply that any individual arbitrageur is well informed, but rather that $y$ is ``known'' collectively. 
Consider sophisticated arbitrageurs who distinguish their interventions $a^{(i)}$, aimed at eliminating return predictability associated with order-flow autocorrelations, from trades $b^{(i)}$ driven by other, independent signals, possibly including fundamental ones. Assume further that the collective arbitrage sector drives $y$ through its aggregate trades $\sum_i b^{(i)}$. From this we cannot yet conclude that $y$ is collectively observed.

The external trader may incidentally distort $b^{(i)}$. Symmetric liquidity effects that remain orthogonal to future returns complicate the analysis but do not alter the covariance argument which leads to white impact returns. A plausible \emph{directional} distortion may arise if liquidity is distorted asymmetrically by $j_t$. In this case, asymmetric changes in the execution conditions of $b^{(i)}$, mediated by $j_t$, may create correlations between $b^{(i)}$ and $j$. However, this is plausibly a second-order effect in $j$.

More seriously, consider the distortion of a fundamental mispricing by the exogenous trader's cumulative impact \emph{state}, $I_t=\sum_{v<t}j_v$. Relative to the counterfactual, a mispricing is shifted by $-\delta^{(i)}I_t$ for some $\delta^{(i)}$. Suppose now that at least one of the arbitrageurs trades on $s_t^{(i)}=y_t^{(i)}-\delta^{(i)}I_t$. When $N=M$, the history of the quantity
\bee
r_t:=x_t-\sum_i s_t^{(i)}=j_t+dI_t \,, \qquad d=\sum_i\delta^{(i)} \,,
\eee
rather than of $y_t$ is collectively observable. At an arbitraged cluster point, orthogonality of returns to past returns and private signals therefore implies $\operatorname{Cov}(x_t,r_u)=0$ for $u<t$; causality of $r_u$ and independence of future fundamental innovations then give
\bee
\operatorname{Cov}(j_t,j_u)+d\sum_{v<u}\operatorname{Cov}(j_t,j_v)=0 \,.
\eee
With $I_0=0$, induction over $u$ yields $\operatorname{Cov}(j_t,j_u)=0$ for every $u<t$. Hence, linear feedback through the impact state preserves impact whiteness.

We conclude that, while aggregate informational completeness is not immediately compelling, there are plausible \emph{a priori} reasons to believe in it. The current framework's reproduction of the known empirical impact phenomenology provides some indication that markets indeed operate in a regime close to informational completeness. Further research is certainly needed to fully examine the interaction between impact and information.

\section{Conclusion}
\label{sec:conclusion}

We have studied how decentralized arbitrage can generate diffusive impact without identifying the underlying trader. On a finite observation window, every accumulation point of the adjustment process has uncorrelated returns; aggregate informational completeness additionally whitens the trader's impact returns. Under the stated stationary-limit assumptions, impact therefore takes a causal all-pass form. Its remaining phase freedom determines trading costs and relaxation. A minimum-distortion criterion yields positive costs and, in a biexponential benchmark, a narrow range of conditional impact decay. Deriving phase selection from trading behavior, and testing the predicted dependence on participant-level order-flow statistics remain the central next steps.

\section*{Disclosure of Interests}

There are no interests to disclose.

\section*{Declaration of Funding}

No funding was received.

\section*{Generative AI statement}

OpenAI's Astra was used during the preparation of this manuscript to assist with proofreading. Furthermore, the candidate family of deformations $V_\tau$ was identified with the assistance of Astra and subsequently verified analytically. Astra also assisted in establishing a rigorous argument about the degree of the all-pass filter under Kronecker's theorem.

\section*{Acknowledgements}

The author acknowledges stimulating discussions with Matteo Marsili, Jean-Philippe Bouchaud, Iacopo Mastromatteo and Galin Georgiev.

\bibliographystyle{plainnat}
\bibliography{square-root-impact}

\end{document}